# Designing Planar, Ultra-Thin, Broad-Band and Material-Versatile Solar Absorbers via Bound-Electron and Exciton Absorption

Dong Liu[†,*], Lin Wang[†]

MIIT Key Laboratory of Thermal Control of Electronic Equipment, School of Energy and Power Engineering, Nanjing University of Science and Technology, Nanjing 210094, China.

[†]These authors contributed equally to this work.

[*]Email address: <u>liudong15@njust.edu.cn</u>.

**Abstract**

Ultrathin planar absorbing layers, including semiconductor and metal films, and 2D materials, are promising building blocks for solar energy harvesting devices but poor light absorption has been a critical issue. Although interference in ultrathin absorbing layers has been studied to realize near perfect absorption at a specific wavelength, achieving high broadband absorption still remains challenging. Here, we both theoretically and experimentally demonstrated a method to tune not only reflection phase shift but also electromagnetic energy dissipation to design broadband solar absorber with simple planar structure consisting of an ultrathin absorbing layer separated from the metallic substrate by a transparent layer. We explicitly identified by deriving a new formulism that the absorbing material with refractive index proportional to the wavelength as well as extinction coefficient independent of the wavelength, is the ideal building block to create ultrathin planar broadband absorbers. To demonstrate the general strategy for naturally available absorbing materials in both high-loss (refractory metals) and weak-absorption (2D materials) regimes, we leveraged the bound-electron interband transition with a broad Lorentz oscillator peak to design a solar thermal absorber based on a ultrathin Cr layer; and leveraged the strong exciton attributed to the spin-orbit coupling for the spectrum near the band edge, and the bound-electron interband transition for shorter wavelengths, to design a solar photovoltaic absorber based on a atomically thin $MoS_2$ layer. Furthermore, our designed ultrathin broadband solar absorbers with planar structures have comparable absorption properties compared to the absorbers with nanopatterns. Our proposed design strategies pave the way to novel nanometer thick energy harvesting and optoelectronic devices with simple planar structures.

# 1. Introduction

Nanometer thick planar films (semiconductor films [1], and two-dimensional (2D) materials [2]) are promising building blocks for solar energy harvesting devices [3] because they can increase device performance and reduce material and fabrication costs [4]. For example, optical absorbers with ultrathin active absorbing layers are attractive for solar photovoltaics because the internal quantum efficiency increases with decreasing active absorbing layer thickness [5]; they also hold promise for solar thermal systems due to the heat localization effect [6]. Furthermore, absorbers with ultrathin thicknesses and ultralight weights have potential for portable and flexible devices [7]. However, the light-matter interaction is weak and the resultant optical absorption is poor in absorbers thinned down to the ultrathin limit. The non-trivial interference supported in ultrathin semiconductor layers has been studied and leveraged to enhance the optical absorption of the absorbers with nanometer thick absorbing layers in many studies [8-12]. These studies focused on designing the reflection phase shift behavior at the semiconductor layer bottom to achieve destructive interference. For example, Song et al. [10] added a transparent layer ($Al_2O_3$) between the semiconductor layer (Ge) and the metallic substrate (Al) with the reflection phase shifts at the Ge layer bottom tuned by the $Al_2O_3$ thicknesses; absorption higher than 80% was achieved in the absorber with 1.5 nm Ge layer for wavelengths below 500 nm. However, two issues arise for this design method. First, narrowband absorption was demonstrated but broadband absorption is more useful for solar absorbers, so to achieve high broadband absorption in ultrathin planar films still remains challenging. Second, medium-loss materials (for example, semiconductors for wavelengths well below the bandgap wavelength) were used for the absorbing layer and this is also one of the reasons why achieving broadband absorption is challenging. Therefore, we are motivated to design ultrathin planar absorbers with broadband absorption and material versatility properties.

Here, we both theoretically and experimentally demonstrated a method to tune not only reflection phase shift but also electromagnetic energy dissipation to design ultrathin planar broadband solar absorbers. The absorber geometry is particularly simple with an ultrathin absorbing material layer separated from the metallic substrate by a transparent material layer. We explicitly identified that the absorbing materials with refractive indices proportional to the wavelength (anomalous dispersion) as well as extinction coefficients independent of the wavelength, are the ideal building block to create ultrathin planar broadband absorbers. Then, we presented two general design strategies for naturally available absorbing materials in both high-loss (refractory metals) and weak-absorption (2D materials) regimes by exploiting their bound-electron and exciton absorption features [13, 14]. Our designed ultrathin solar thermal and photovoltaic absorbers with simple planar structures have comparable absorption properties compared to the absorbers with nanophotonic structures that require sophisticated lithography steps [4, 15-18].

## 2. Results and discussion
### 2.1 Proof of concept

In this section, we explicitly identified the ideal complex refractive index of the absorbing material for the absorber structure shown in Figure 1 to achieve perfect broadband absorption in the ultrathin layer based on this absorbing material. Two assumptions were made: (i) the magnitude of the complex refractive index of the absorbing material is much larger than one; (ii) the metallic substrate is approximated as perfect electric conductor. The first assumption is reasonable for most semiconductors and metals. Regarding the second assumption, no restriction was placed on the metallic substrate (even metals approaching the perfect electric conductor limit can be used) in this work in contrast to Kats et al. [8] and Rensberg et al. [19] where metals with finite optical conductivity or near zero permittivity were used.

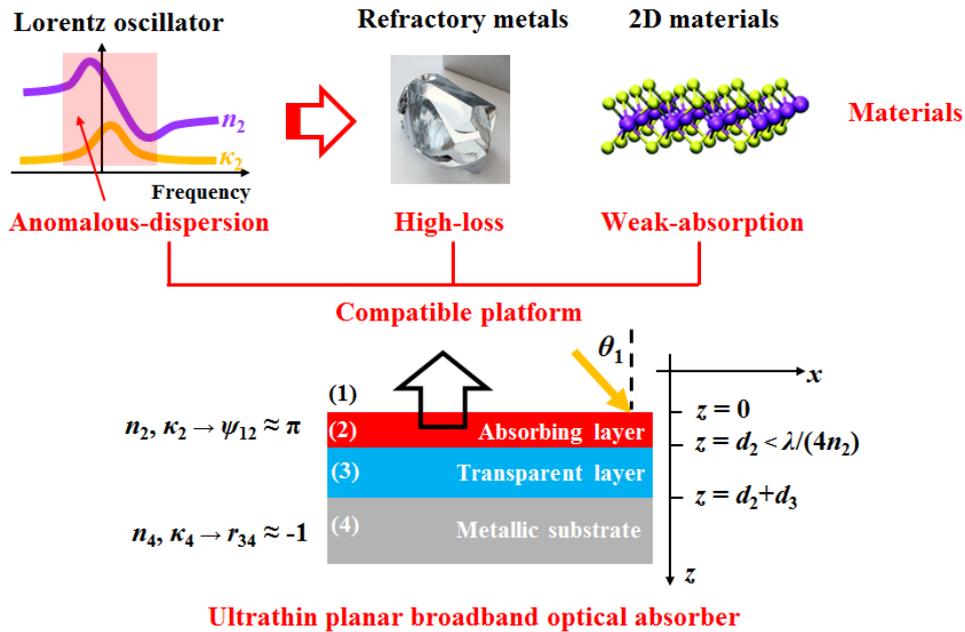

**Figure 1. Schematic of an ultrathin planar broadband absorber.** The absorber consists of an ultrathin thin absorbing material layer (refractive index $n_2$, extinction coefficient $\kappa_2$, thickness $d_2$), a transparent material layer (thickness $d_3$) and an opaque metallic substrate. This absorber is a compatible platform for absorbing materials in both the high-loss (for example, refractory metals) and the weak-absorption (for example, 2D materials) regimes.

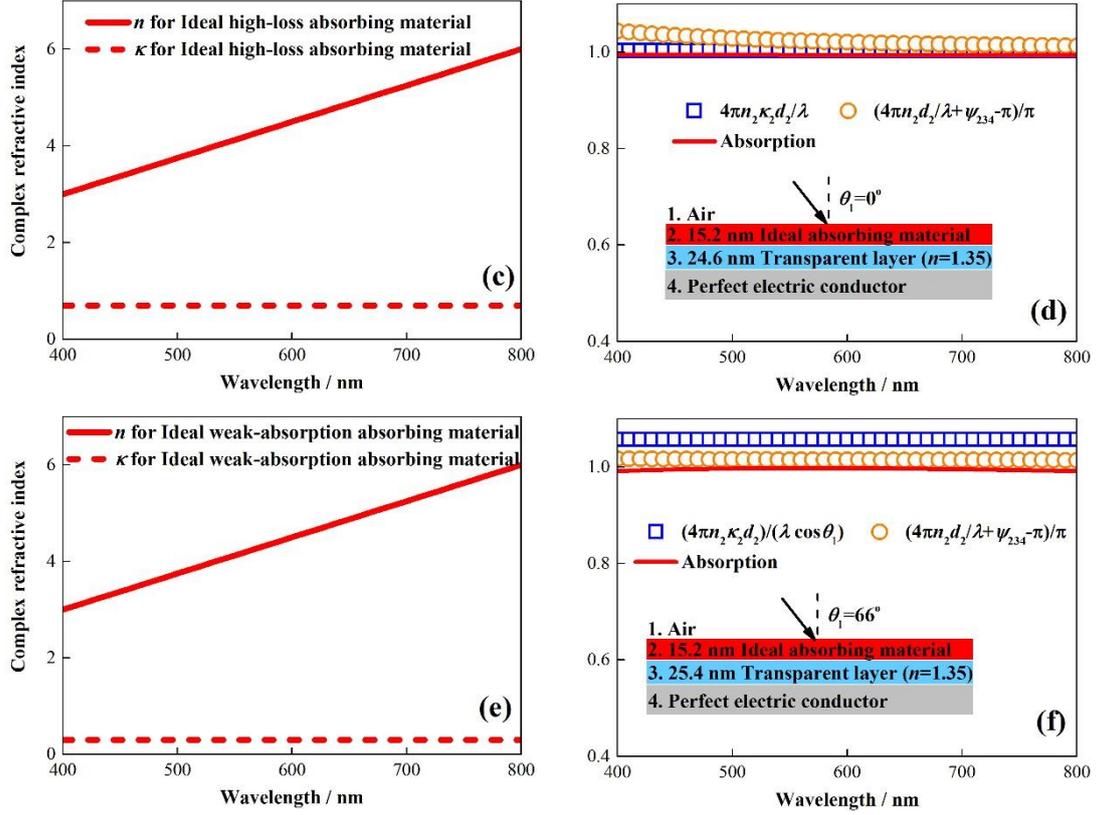

**Figure 2. Designing ultrathin planar broadband absorbers.** Complex refractive indices of (c) the ideal high-loss and (e) the ideal weak-absorption absorbing materials. Calculated absorptivity, $4\pi n_2 \kappa_2 d_2/(\lambda\cos\theta_1)$ and $(4\pi n_2 d_2/\lambda+\psi_{234}-\pi)/\pi$ of the absorber consisting of (d) 15.2 nm ideal high-loss absorbing material, 24.6 nm transparent material ($n_3$=1.35) and a perfect electric conductor substrate for normal incidence, and (f) 15.2 nm ideal weak-absorption absorbing material, 25.4 nm transparent material ($n_3$=1.35) and a perfect electric conductor substrate for 66° incidence, demonstrating near perfect broadband absorption.

We designed an ultrathin absorber with perfect broadband absorption. We can explicitly derive the ideal complex refractive index of the absorbing material as

$$\frac{n_2(\lambda)}{\lambda} = \text{constant} \tag{14}$$

$$\kappa_2(\lambda) = \text{constant} > 0.64 \tag{15}$$

to achieve perfect broadband absorption in the ultrathin layer based on this absorbing material. To validate Equations (14) and (15), we "produced" an ideal absorbing material with $n_2(\lambda) = 3\lambda/400$ and $\kappa_2(\lambda) = 0.7$ as shown in Figure 2c and "produced" the absorber with an ultrathin ideal absorbing material layer separated from the perfect electric conductor substrate by a transparent material layer (assuming $n_3 = 1.35$ close to the refractive index of MgF$_2$ without losing the generality) as shown in Figure 2d. The thickness of the ideal absorbing material is designed to be 15.2 nm with the thickness of the transparent material layer designed to be 24.6 nm. We investigated the absorption property of this absorber for wavelengths from 400 to 800 nm. Figure 2d shows that the calculated $4\pi n_2 \kappa_2 d_2/\lambda$ and $(4\pi n_2 d_2/\lambda + \psi_{234} - \pi)/\pi$ are both close to one over the entire spectrum. Results also show that the calculated absorptivity is higher than 99% over the entire spectrum demonstrating near perfect broadband absorption behavior and validating Equations (14) and (15). Therefore, the material with refractive index proportional to the wavelength, and with the extinction coefficient independent of the wavelength and larger than 0.64, is the ideal building block to create ultrathin absorber with perfect broadband absorption.

In the next step, we investigated the weak absorption cases. One case is that the extinction coefficient of the absorbing material is smaller than 0.64. The other case is that the thickness of the absorbing material layer is small. These two cases are equivalent and can be generalized to the weak-absorption regime expressed as

$$\frac{4\pi n_2 \kappa_2 d_2}{\lambda} \ll 1 \tag{16}$$

We consider the oblique incident TE polarized light (our proposed method can be extended to transverse-magnetic, TM, polarization, see Section 2 of the Supplementary material for details). The absorption rate per unit length for oblique incident condition is expressed as (see Section 1 of the Supplementary material for details)

$$A_{pul}(z,\lambda) = \frac{4\pi n_2 \kappa_2}{\lambda \cos\theta_1} \frac{|E(z)|^2}{|E_0|^2} \quad (17)$$

where $\theta_1$ is the incident angle, so Equation (13) turns into

$$\frac{4\pi n_2 \kappa_2 d_2}{\lambda \cos\theta_1} = 1 \quad (18)$$

Since $\cos\theta_1 < 1$, there is always a $\theta_1$ at which Equation (18) is satisfied for materials in the weak-absorption regime. Equation (15) resultantly turns into

$$\kappa_2(\lambda) = \text{constant} \quad (19)$$

Therefore, Equation (18) are the more general guidelines to design ultrathin planar broadband absorbers. To validate this, we "produced" an ideal absorbing material with $n_2(\lambda) = 3\lambda/400$ and $\kappa_2(\lambda) = 0.3 < 0.64$ as shown in Figure 2e and "produced" the absorber with an ultrathin ideal absorbing material layer separated from the perfect electric conductor substrate by a transparent material layer ($n_3 = 1.35$) as shown in Figure 2f. The thickness of the ideal absorbing material is fixed at 15.2 nm as same as the thickness for the absorber in Figure 2d. The incident angle is designed to be 66° according to Equation (18). The thickness of the transparent material layer is then designed to be 25.4 nm according to Equation (8). Figure 2f shows that the calculated $4\pi n_2 \kappa_2 d_2 / (\lambda \cos\theta_1)$ and $(4\pi n_2 d_2 / \lambda + \psi_{234} - \pi)/\pi$ are both close to one from 400 to 800 nm wavelengths corresponding to Equation (18). Results also show that the calculated absorptivity is higher than 99% over the entire visible spectrum validating Equations (14) and (19).

So far, we have demonstrated that designing ultrathin planar absorber with near perfect broadband absorption can be generally guided by Equation (18). The ideal absorbing material should have refractive index proportional to the wavelength and extinction coefficient independent of the wavelength. There are no other constraints on either the extinction coefficient magnitude or on the thickness, so our designed absorber is a compatible platform for high-loss, low-loss and atomically thin materials. In following sections, we designed ultrathin planar broadband solar absorbers based on naturally available materials in both high-loss and weak-absorption regimes.

**2.2 High-loss regime**

Equations (14) and (19) features anomalous dispersion that can be modeled as a Lorentz oscillator. Semiconductors have this anomalous dispersion for the spectrum near the peak wavelength of the Lorentz oscillator due to the bound-electron interband transition [14]. However, the Lorentz oscillator peak wavelengths for semiconductors are usually located in the ultraviolet spectrum or in the short wavelength region of the visible spectrum. In this section, we leverage the interband transition of the refractory metals whose relatively broad Lorentz oscillator peak wavelengths can be located in visible and near infrared spectra, which can be intuitively understood by considering them as zero-bandgap semiconductors [22].

Refractory metals usually have large extinction coefficients and resultant high reflectivity in visible and near infrared spectra due to the significant impedance mismatch at the interfaces with air. For example, Figure 1 shows the reflective appearance of Chromium (Cr, its extinction coefficient is around four in the visible spectrum). Here, we show that designed by our derived formulism, ultrathin planar Cr films achieved high broadband absorption which is counter-intuitive and promising for solar thermal absorbers.

Figure 3a shows the complex refractive index of Cr [23]. We can observe that the variations of its refractive index and extinction coefficient with the wavelength from 600 to 1000 nm are close to the relationships described by Equations (14) and (19). We then designed the absorber with a 4.3 nm Cr layer, a 116.6 nm MgF$_2$ layer and an Au substrate as shown in Figure 3b. Results in Figure 3c show that the calculated absorptivity is higher than 95% from 600 to 1000 nm. This near perfect broadband absorption property was achieved by the Cr film with thickness to be only 1/40 of the wavelength. We also calculated $4\pi n_2 \kappa_2 d_2 / \lambda$ and $(4\pi n_2 d_2 / \lambda + \psi_{234} - \pi)/\pi$ of this absorber. They are close to one over the entire spectrum agreeing well with the theoretical investigation described by Equation (18) and demonstrating the effectiveness of our proposed design method.

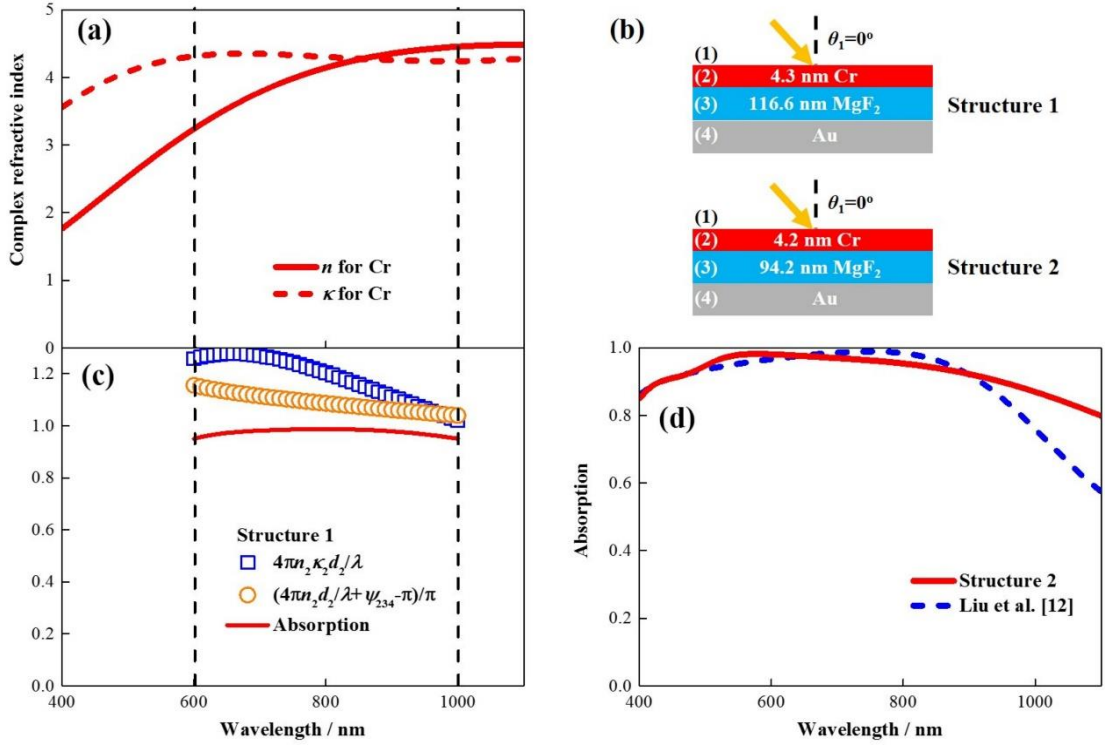

**Figure 3. Ultrathin planar broadband solar thermal absorbers using absorbing materials in the high-loss regime.** (a) Complex refractive indices of Cr. (b) Designed two absorber structures. Structure 1 consists of 4.3 nm Cr, 116.6 nm $MgF_2$ and an opaque Au layer. Structure 2 consists of 4.2 nm Cr, 94.2 nm $MgF_2$ and an opaque Au layer. (c) Calculated absorptivity, $4\pi n_2\kappa_2 d_2/\lambda$ and $(4\pi n_2 d_2/\lambda+\psi_{234}-\pi)/\pi$ of structure 1, demonstrating near perfect broadband absorption from 600 to 1000 nm. (d) Absorptivity of structure 2 and the absorber in Liu et al. [12], demonstrating high broadband absorption from 400 to 1100 nm.

To compare with broadband absorbers in the literature, we designed an absorber with the 4.2 nm Cr/94.2 nm $MgF_2$/Au structure as shown in Figure 3b. Its calculated spectral absorptivity is shown in Figure 3d. The state-of-the-art ultrathin planar broadband absorber reported in Liu et al. [12] has absorptivity higher than 80% from 400 to 950 nm. We can observe that our designed absorber has much broader high absorption band with absorptivity higher than 85% from 400 to 1000 nm and higher than 80% from 400 to 1100 nm. In addition, the absorbing material layer thickness is only 1/3 of that in Liu et al. [12], which is useful for wearable and flexible solar thermal absorbers [7]. This absorber with simple planar structure also has superior broadband absorption property compared to the absorber employing complex nanophotonic structure with absorptivity higher than 85% from 400 to 800 nm and higher than 80% from 400 to 900 nm reported in Li et al. [16]. Furthermore, our designed absorber has a solar spectrum averaged absorptivity of 96% in the visible spectrum which is highly desirable for efficient solar thermal absorbers.

**2.3 Weak-absorption regime**

Mono-layer or few-layer 2D materials and semiconductors for the wavelengths near the band edge are in this regime. 2D materials have shown unique optoelectronic properties for solar cell applications [24]; and few-layer 2D materials have intrinsic layered crystallographic structures in contrast to conventional photovoltaic materials that suffer the degradation of crystalline quality and the resultant low photovoltaic efficiency when thinned down to the ultrathin limit [5, 24]. In the spectrum near the semiconductor band edge, solar cells have higher internal quantum efficiency [25], lower energy loss to lattice vibration, and resultant higher power conversion efficiency. Therefore, it is of interest to explore the possibility to achieve high broadband absorption in the weak-absorption regime.

In this section, we designed the ultrathin planar broadband solar photovoltaic absorber leveraged the strong excitons that have been reported in mono-layer and few-layer transition metal dichalcogenides, black phosphorus and organic-inorganic perovskites [13, 26, 27], and that can also be modeled as Lorentz oscillators [28]. Four-layer MoS$_2$ were used as the representative material with excitons attributed to the splitting of the valence band by the spin-orbit coupling [28]. Here, we focused on optical absorption design (a preliminarily viable solar cell structure was shown in Section 3 of the Supplementary material).

Figure 4a shows the spectral complex refractive index of monolayer MoS$_2$ [28]. We can observe that it features anomalous dispersion for the wavelengths from 580 nm to the bandgap wavelength with the indices for 580 nm, 630 nm and 670 nm approximately satisfied

$$\frac{n_2(\lambda_L = 580 \text{ nm})}{\lambda_L} \approx \frac{n_2(\lambda_M = 630 \text{ nm})}{\lambda_M} \approx \frac{n_2(\lambda_H = 670 \text{ nm})}{\lambda_H} \quad (20)$$

$$\kappa_2(\lambda_L = 580 \text{ nm}) \approx \kappa_2(\lambda_M = 630 \text{ nm}) \approx \kappa_2(\lambda_H = 670 \text{ nm}) \quad (21)$$

corresponding to Equations (14) and (19). We then designed the absorber consisting of a MoS$_2$ layer, a 79 nm MgF$_2$ layer and an Ag substrate for 66 °incident angle as shown in Figure 4b. Results in Figure 4c show that calculated absorptivity is higher than 99% for 580 nm, 630 nm and 670 nm wavelengths and calculated $4\pi n_2 \kappa_2 d_2 / (\lambda \cos\theta_1)$ and $(4\pi n_2 d_2 / \lambda + \psi_{234} - \pi)/\pi$ are close to one for these targeted wavelengths agreeing well with theoretical investigation described by Equation (18).

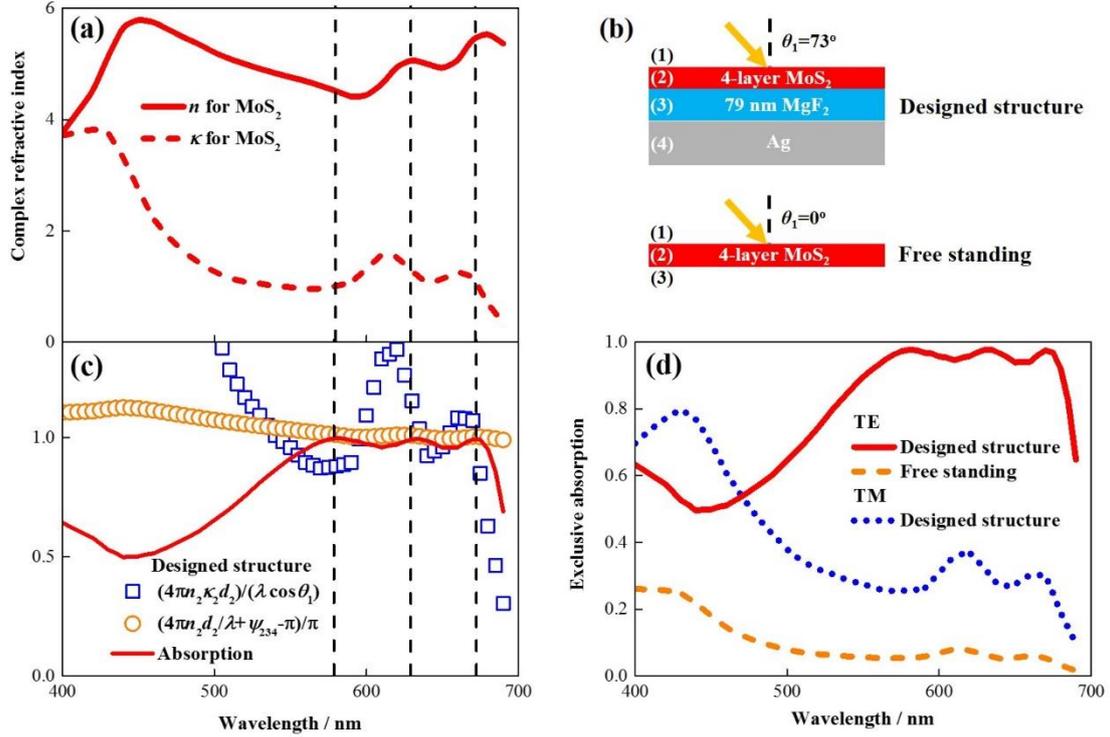

**Figure 4. Ultrathin planar broadband solar photovoltaic absorbers using absorbing materials in the weak-absorption regime.** (a) Complex refractive indices of $MoS_2$. (b) Designed absorber structure consisting of a $MoS_2$ layer, a 79 nm $MgF_2$ layer and an opaque Ag layer, and the free standing structure. (c) Calculated absorptivity, $4\pi n_2 \kappa_2 d_2/(\lambda \cos\theta_1)$ and $(4\pi n_2 d_2/\lambda+\psi_{234}-\pi)/\pi$ of the designed structure, demonstrating high broadband absorption over the entire spectrum below the band edge. (d) Calculated exclusive absorption in the $MoS_2$ layer of the designed and free standing structures for TE polarization, and of the designed structure for TM polarization.

We compared the exclusive absorption in the $MoS_2$ layer between our designed structure and the free standing structure in Figure 4d. Results shown that attributed to the exciton Lorentz oscillator peak (much broader than the interband transition Lorentz oscillator peak from 400 to 450 nm as shown in Figure 4a), our designed absorber has broadband absorption higher than 80% (one order of magnitude enhancement of absorption compared to the free standing structure) from 530 nm to the bandgap wavelength where the photovoltaic internal quantum efficiency is high. Results also show that for the wavelengths below 530 nm, the absorption of our designed absorber has a two-fold enhancement compared to the free standing structure due to the round-trip intrinsic interband transition absorption. These results demonstrated considerable broadband absorption of our designed absorber given its atomic-layer thickness. Therefore, the general strategy to design ultrathin planar broadband solar photovoltaic absorber is to leverage the strong exciton for the wavelengths near the band edge and to leverage the bound-electron interband transition for shorter wavelengths.

To compare with the absorber employing nanopatterns in Huang et al. [17], we calculated the photocurrent ($I_{ph}$, mA cm$^{-2}$) expressed as

$$I_{ph} = e \int_{400 \text{ nm}}^{690 \text{ nm}} \frac{A_{exc, MoS_2}(\lambda) S(\lambda)}{hc/\lambda} \quad (22)$$

where $e$ is the unit charge, $S$ is the ASTM Air Mass 1.5 solar irradiance, $h$ is Planck's constant, $c$ is the speed of light and $A_{exc, MoS_2}$ is exclusive absorption in the MoS$_2$ layer. The photocurrent of our designed absorber is 13.1 mA cm$^{-2}$ comparable to 13.5 mA cm$^{-2}$ in Huang et al. [17]. In addition, we calculated the exclusive absorption in the MoS$_2$ layer of our designed absorber for TM polarization as shown in Figure 4d. Variations higher than 50% in absorption for TE and TM polarizations are shown for wavelengths from 510 nm to the bandgap wavelength demonstrating the broadband polarization-selective absorption behavior. Therefore, our proposed design method also paves the way to photodetection devices with new functionalities because the extra polarization information is detected from optical signals [29, 30].

## 3. Conclusions

Finally, we summarize two general strategies to design ultrathin planar broadband solar absorbers. In the high-loss regime, the broad Lorentz oscillator peak of the refractory metal attributed to the bound-electron interband transition can be leveraged to achieve high broadband absorption. We demonstrated a solar thermal absorber consisting of an only 4.2 nm thick Cr layer with absorptivity higher than 85% from 400 to 1000 nm and higher than 80% from 400 to 1100 nm, and with a solar spectrum averaged absorptivity of 96% in the visible spectrum. In the weak-absorption regime, the exciton of 2D materials for wavelengths near the bandgap wavelength combining with the interband transition for wavelengths well below the bandgap wavelength, can be leveraged to enhance optical absorption over a wide spectrum. We demonstrated a solar photovoltaic absorber consisting of an atomically thin $MoS_2$ layer with one order of magnitude enhancement of absorption from 530 nm to the bandgap wavelength, and with a two-fold enhancement below 530 nm, compared to the free standing structure.

## 4. Methods

The expression of the absorption rate per unit length normalized to the incident energy was derived based on the transfer matrix theory. This theory was also used to calculate the total absorption (or the reflectivity) and the exclusive absorption in the absorbing material layers, shown in Figures 2 to 4 (see Section 1 of the Supplementary material for details).

**Acknowledgments**

This work was supported by the National Natural Science Foundation of China

**References**

[1] V. Steenhoff, M. Juilfs, R.E. Ravekes, M. Vehse, C. Agert, Resonant-cavity-enhanced a-Ge:H nanoabsorber solar cells for application in multijunction devices. Nano Energy 27 (2016) 658-663.

[2] X. Yu, S. Zhang, H. Zeng, Q.J. Wang, Lateral black phosphorene p-n junctions formed via chemical doping for high performance near-infrared photodetector. Nano Energy 25 (2016) 34-41.

[3] H. Li, Y. Shi, M.H. Chiu, L.J. Li, Emerging energy applications of two-dimensional layered transition metal dichalcogenides. Nano Energy 18 (2015) 293-305.

[4] S.J. Kim, P. Fan, J.H. Kang, M.L. Brongersma, Creating semiconductor metafilms with designer absorption spectra. Nat. Commun. 6 (2015) 7591.

[5] Z. Xia, H. Song, M. Kim, M. Zhou, T.H. Chang, D. Liu, X. Yin, K. Xiong, H. Mi, X. Wang, F. Xia, Z. Yu, Z. Ma, Q. Gan, Single-crystalline germanium nanomembrane photodetectors on foreign nanocavities. Sci. Adv. 3 (2017) e1602783.

[6] H. Ghasemi, G. Ni, A.M. Marconnet, J. Loomis, S. Yerci, N. Miljkovic, G. Chen, Solar steam generation by heat localization. Nat. Commun. 5 (2014) 4449.

[7] Y.S. Jung, D.H. Jeong, S.B. Kang, F. Kim, M.H. Jeong, K.S. Lee, J.S. Son, J.M.

Baik, J.S. Kim, K.J. Choi, Wearable solar thermoelectric generator driven by unprecedentedly high temperature difference. Nano Energy 40 (2017) 663-672.

[8] M.A. Kats, R. Blanchard, P. Genevet, F. Capasso, Nanometre Optical Coatings Based on Strong Interference Effects in Highly Absorbing Media. Nat. Mater. 12 (2013) 20-24.

[9] K.T. Lee, S. Seo, J.Y. Lee, L.J. Guo, Strong resonance effect in a lossy medium-based optical cavity for angle robust spectrum filters. Adv. Mater. 26 (2014) 6324-6328.

[10] H. Song, L. Guo, Z. Liu, K. Liu, X. Zeng, D. Ji, N. Zhang, H. Hu, S. Jiang, Q. Gan, Nanocavity enhancement for ultra-thin film optical absorber. Adv. Mater. 26 (2014) 2737-2743.

[11] D. Liu, Q. Li, Sub-nanometer planar solar absorber. Nano Energy 34 (2017) 172-180.

[12] D. Liu, H.T. Yu, Z. Yang, Y.Y. Duan, Ultrathin planar broadband absorber through effective medium design. Nano Res. 9 (2016) 2354-2363.

[13] T. Low, A. Chaves, J.D. Caldwell, A. Kumar, N.X. Fang, P. Avouris, T.F. Heinz, F. Guinea, L. Martin-Moreno, F. Koppens, Polaritons in layered two-dimensional materials. Nat. Mater. 16 (2017) 182-194.

[14] C.F. Bohren, D.R. Huffman, Absorption and Scattering of Light by Small Particles, Wiley, New York, 1983.

[15] C. Ji, K.T. Lee, T. Xu, J. Zhou, H.J. Park, L.J. Guo, Engineering light at the nanoscale: structural color filters and broadband perfect absorbers. Adv. Opt.

Mater. DOI: 10.1002/adom.201700368.

[16] W. Li, U. Guler, N. Kinsey, G.V. Naik, A. Boltasseva, J. Guan, V.M. Shalaev, A.V. Kildishev, Refractory plasmonics with titanium nitride: broadband metamaterial absorber. Adv. Mater. 26 (2014) 7959-7965.

[17] L. Huang, G. Li, A. Gurarslan, Y. Yu, R. Kirste, W. Guo, J. Zhao, R. Collazo, Z. Sitar, G.N. Parsons, M. Kudenov, L. Cao, Atomically thin $MoS_2$ narrowband and broadband light superabsorbers. ACS Nano 10 (2016) 7493-7499.

[18] Z. Wu, K. Chen, R. Menz, T. Nagao, Y. Zheng, Tunable multiband metasurfaces by moire nanosphere lithography. Nanoscale 7 (2015) 20391-20396.

[19] J. Rensberg, Y. Zhou, S. Richter, C. Wan, S. Zhang, P. Schoppe, R. Schmidt-Grund, S. Ramanathan, F. Capasso, M.A. Kats, C. Ronning, Epsilon-near-zero substrate engineering for ultrathin-film perfect absorbers. Phys. Rev. Appl. 8 (2017) 014009.

[20] D. Liu, H.T. Yu, Y.Y. Duan, Q. Li, Y.M. Xuan, New insight into the angle insensitivity of ultrathin planar optical absorbers for broadband solar energy harvesting. Sci. Rep. 6 (2016) 32515.

[21] J. Park, J.H. Kang, A.P. Vasudev, D.T. Schoen, H. Kim, E. Hasman, M.L. Brongersma, Omnidirectional near-unity absorption in an ultrathin planar semiconductor layer on a metal substrate. ACS Photon. 1 (2014) 812-821.

[22] E.D. Palik, Handbook of Optical Constants of Solids, Academic Press, London, 1997.

[23] A.D. Rakic, A.B. Djurisic, J.M. Elazar, M.L. Majewski, Optical properties of metallic films for vertical-cavity optoelectronic devices. Appl. Opt. 37 (1998)

5271-5283.

[24] J. Wong, D. Jariwala, G. Tagliabue, K. Tat, A.R. Davoyan, M.C. Sherrott, H.A. Atwater, High photovoltaic quantum efficiency in ultrathin van der Waals heterostructures. ACS Nano 11 (2017) 7230-7240.

[25] A. Lenert, D.M. Bierman, Y. Nam, W.R. Chan, I. Celanovic, M. Soljacic, E.N. Wang, A nanophotonic solar thermophotovoltaic device. Nat. Nanotech. 9 (2014) 126-130.

[26] J. Zhang, T. Wu, J. Duan, M. Ahmadi, F. Jiang, Y. Zhou, B. Hu, Exploring spin-orbital coupling effects on photovoltaic actions in Sn and Pb based perovskite solar cells. Nano Energy 38 (2017) 297-303.

[27] A. Molina-Sanchez, D. Sangalli, K. Hummer, A. Marini, L. Wirtz, Effect of spin-orbit interaction on the optical spectra of single-layer, double-layer, and bulk $MoS_2$. Phys. Rev. B 88 (2013) 045412.

[28] Y. Li, A. Chernikov, X. Zhang, A. Rigosi, H.M. Hill, A.M. van der Zande, D.A. Chenet, E.M. Shih, J. Hone, T.F. Heinz, Measurement of the optical dielectric function of monolayer transition-metal dichalcogenides: $MoS_2$, $MoSe_2$, $WS_2$, and $WSe_2$. Phys. Rev. B 90 (2014) 205422.

[29] H. Yuan, X. Liu, F. Afshinmanesh, W. Li, G. Xu, J. Sun, B. Lian, A.G. Curto, G. Ye, Y. Hikita, Z. Shen, S.C. Zhang, X. Chen, M. Brongersma, H.Y. Hwang, Y. Cui, Polarization-sensitive broadband photodetector using a black phosphorus vertical p-n junction. Nat. Nanotechnol. 10 (2015) 707-713.

[30] L. Ye, P. Wang, W. Luo, F. Gong, L. Liao, T. Liu, L. Tong, J. Zang, J. Xu, W. Hu,

Highly polarization sensitive infrared photodetector based on black phosphorus-on-WSe2 photogate vertical heterostructure. Nano Energy 37 (2017) 53-60.

**Designing Planar, Ultra-Thin, Broad-Band and Material-Versatile**

**Solar Absorbers via Bound-Electron and Exciton Absorption**

Dong Liu[†,*], Lin Wang[†]

MIIT Key Laboratory of Thermal Control of Electronic Equipment, School of Energy and Power Engineering, Nanjing University of Science and Technology, Nanjing 210094, China.

[†]These authors contributed equally to this work.

[*]Email address: liudong15@njust.edu.cn.

## 1. Transfer matrix theory

Here, we derived the expression of the absorption rate per unit length normalized to the incident energy in a specified layer of a multilayer structure. The electric and magnetic fields, and the wave vector are the superposition of the electromagnetic waves heading forward and backward. The wave vector is in the *x-z* plane shown in Figure 1. For TE polarization, we have

$$\boldsymbol{E}_\text{f}^\text{TE} = E_\text{f}\boldsymbol{y},\ \boldsymbol{E}_\text{b}^\text{TE} = E_\text{b}\boldsymbol{y} \tag{S1}$$

$$\boldsymbol{k}_\text{f} = \frac{2\pi}{\lambda}m\left[\cos(\theta)\boldsymbol{z} + \sin(\theta)\boldsymbol{x}\right],\ \boldsymbol{k}_\text{b} = \frac{2\pi}{\lambda}m\left[-\cos(\theta)\boldsymbol{z} + \sin(\theta)\boldsymbol{x}\right] \tag{S2}$$

$\boldsymbol{H} = (\mu/\omega)\boldsymbol{k}\times\boldsymbol{E}$ gives

$$\boldsymbol{H}_\text{f}^\text{TE} \propto mE_\text{f}\left[-\cos(\theta)\boldsymbol{z} + \sin(\theta)\boldsymbol{x}\right],\ \boldsymbol{H}_\text{b}^\text{TE} \propto mE_\text{b}\left[\cos(\theta)\boldsymbol{z} + \sin(\theta)\boldsymbol{x}\right] \tag{S3}$$

The magnitude of the Poynting vector is expressed as

$$|\boldsymbol{S}|^\text{TE} = \frac{1}{2}\text{Re}\left[\boldsymbol{z}\cdot\left(\overline{\boldsymbol{E}}^\text{TE}\times\boldsymbol{H}^\text{TE}\right)\right] \propto \text{Re}\left[\left(\overline{E}_\text{f} + \overline{E}_\text{b}\right)\left(E_\text{f} - E_\text{b}\right)m\cos(\theta)\right] \tag{S4}$$

where the bar means the conjugate operation. Normalizing to the incident energy gives

$$|S|_{\text{norm}}^{\text{TE}} = \frac{\text{Re}\left[\left(\overline{E}_{\text{f}} + \overline{E}_{\text{b}}\right)\left(E_{\text{f}} - E_{\text{b}}\right)m\cos(\theta)\right]}{\text{Re}\left[m_1\cos(\theta_1)\right]|E_0|^2} \quad (S5)$$

The absorption rate per unit length normalized to the incident energy is equal to the negative derivative of Equation (S5) expressed as

$$A_{\text{pul}}^{\text{TE}}(z,\lambda) = \frac{|E_{\text{f}} + E_{\text{b}}|^2 \text{Im}\left[2\pi m^2 \cos^2(\theta)/\lambda\right]}{|E_0|^2 \text{Re}\left[m_1\cos(\theta_1)\right]} = \frac{4\pi n_2 \kappa_2}{\lambda \cos\theta_1} \frac{|E(z)|^2}{|E_0|^2} \quad (S6)$$

For TM polarization, we have

$$\boldsymbol{E}_{\text{f}}^{\text{TM}} = E_{\text{f}}\left[-\cos(\theta)\boldsymbol{z} + \sin(\theta)\boldsymbol{x}\right], \ \boldsymbol{E}_{\text{b}}^{\text{TM}} = E_{\text{b}}\left[\cos(\theta)\boldsymbol{z} + \sin(\theta)\boldsymbol{x}\right] \quad (S7)$$

$$\boldsymbol{H}_{\text{f}}^{\text{TM}} \propto mE_{\text{f}}\boldsymbol{y}, \ \boldsymbol{H}_{\text{b}}^{\text{TM}} \propto mE_{\text{b}}\boldsymbol{y} \quad (S8)$$

$$A_{\text{pul}}^{\text{TM}}(z,\lambda) \approx \frac{4\pi n_2 \kappa_2}{\lambda \cos\theta_1} \frac{|(E_{\text{f}} - E_{\text{b}})\cos(\theta)|^2}{|E_0|^2} \quad (S9)$$

Details to calculate the total absorption (or the reflectivity), and the exclusive absorption in a specified layer, were given in our previous works [S1, S2].

## 2. Results for TM polarization

First, we applied the transfer matrix theory to the ultrathin absorbing material layer in the structure shown in Figure 1 for TM polarization. The exclusive absorption in the ultrathin absorbing material layer is

$$A_{\text{exc}}^{\text{TM}}(\lambda) = \frac{4\pi n_2 \kappa_2}{\lambda \cos\theta_1} \frac{\int_0^{d_2}|(E_{\text{f}} - E_{\text{b}})(z)\cos(\theta_2)|^2 dz}{|E_0|^2} \quad (S10)$$

We assume that

$$(E_{\text{f}} - E_{\text{b}})(z)\cos(\theta_2) = (E_{\text{f}} - E_{\text{b}})(z=0)\cos(\theta_2) = E_0\cos(\theta_1), \ 0 \leq z \leq d_2 \quad (S11)$$

as we did in the manuscript. By substituting Equation (S11) into Equation (S10) and

applying the perfect absorption condition ($A_{exc}^{TM} = 1$), we get

$$\frac{4\pi n_2 \kappa_2 d_2 (\cos\theta_1)^2}{\lambda \cos\theta_1} = \frac{4\pi n_2 \kappa_2 d_2}{\lambda}\cos\theta_1 = 1 \qquad (S12)$$

By comparing Equation (S12) to Equation (18) we found that the structure in Figure 1 cannot achieve perfect absorption in weak-absorption materials for TM polarization because $\cos\theta_1$ is in the numerator.

Therefore, we then designed the structure (shown in Figure S1a) with a top transparent material layer added to the structure in Figure 1. The perfect absorption scenario for this structure can be approximated to the destructive interference condition between reflected electromagnetic waves at 2–3 and 3–4 interfaces because reflection coefficients at these two interfaces are much larger than that at 1–2 interface. Therefore, the ideal refractive index, $n_{3,ideal}$, extinction coefficient, $\kappa_{3,ideal}$ and thickness of the absorbing material, $d_{3,ideal}$ and the ideal incident angle, $\theta_{1,ideal}$, can be reasonably estimated to satisfy

$$\frac{4\pi n_{3,ideal} \kappa_{3,ideal} d_{3,ideal} (\cos\theta_2)^2}{\lambda \cos\theta_2} < 1 < \frac{4\pi n_{3,ideal} \kappa_{3,ideal} d_{3,ideal} (\cos\theta_2)^2}{\lambda \cos\theta_{1,ideal}} \qquad (S13)$$

where $m_2 \cos\theta_2 = m_1 \cos\theta_{1,ideal}$. Since $m_1 < m_2$, $\cos\theta_1 < \cos\theta_2$; thus, the structure in Figure S1a can achieve perfect absorption based on materials in the weak-absorption regime for TM polarization.

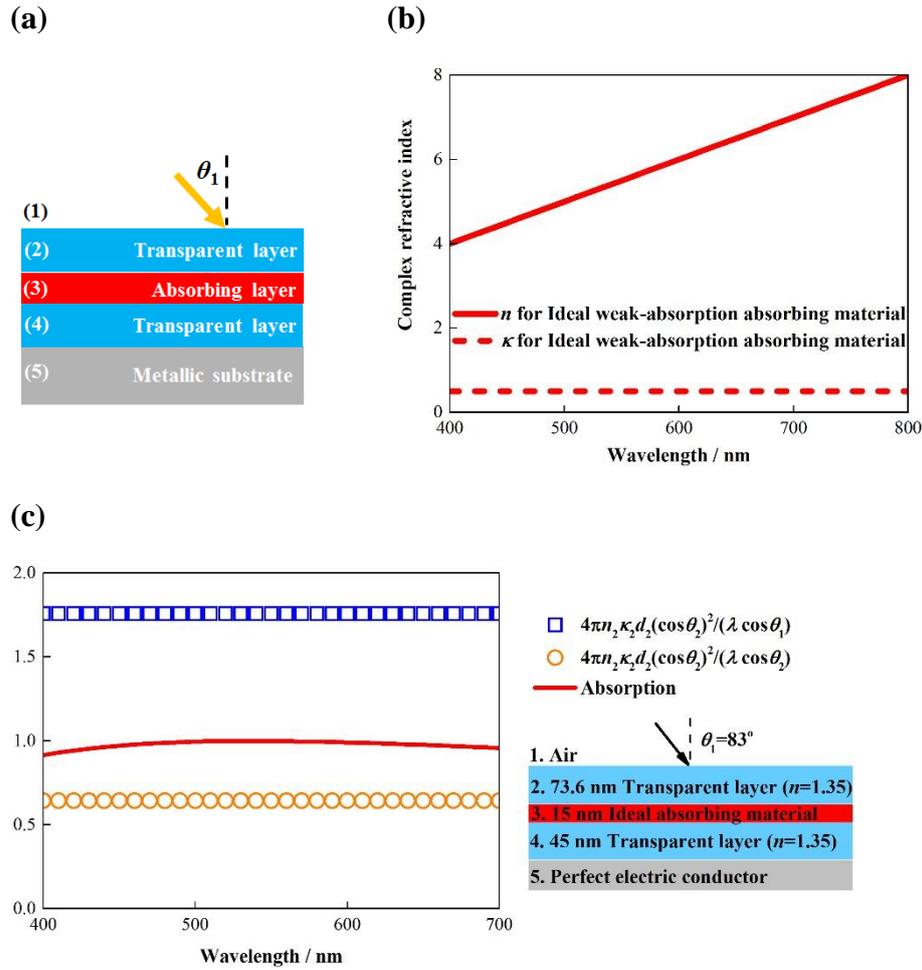

**Figure S1. Designing ultrathin planar broadband absorbers for TM polarization.**

(a) Schematic of the absorber structure with a top transparent material layer added to the structure in Figure 1. (b) Complex refractive index of the ideal weak-absorption absorbing material. (c) Calculated absorptivity, $4\pi n_3 \kappa_3 d_3 (\cos\theta_2)^2 / (\lambda \cos\theta_2)$ and $4\pi n_3 \kappa_3 d_3 (\cos\theta_2)^2 / (\lambda \cos\theta_1)$ of the absorber consisting of a 73.6 nm transparent material top-layer ($n_3 = 1.35$), a 15 nm ideal absorbing material layer, a 45 nm transparent material inter-layer ($n_3 = 1.35$) and a perfect electric conductor substrate for 83 °incident TM polarized light, demonstrating near perfect broadband absorption.

In our previous work [S1], we have demonstrated that there is always a pair of $(\theta_1, d_2)$ for which near perfect absorption at a specified wavelength is achieved by this structure with a MoS$_2$ monolayer (weak-absorption regime). Now we prove that Equations (14) and (19) can also be applied to this structure for TM polarization as the material criteria in the weak-absorption regime to create ultrathin planar broadband absorbers. we "produced" an ideal absorbing material with $n_2(\lambda) = \lambda/100$ and $\kappa_2(\lambda) = 0.5$ for the spectrum from 400 to 800 nm as shown in Figure S1b and "produced" the absorber consisting of a 73.6 nm transparent material top-layer ($n_3 = 1.35$), a 15 nm ideal absorbing material layer, a 45 nm transparent material inter-layer ($n_3 = 1.35$) and a perfect electric conductor substrate for 83° incident TM polarized light as shown in Figure S1c. Results in Figure S1c show that the calculated absorption curve is between the calculated $4\pi n_3 \kappa_3 d_3 (\cos\theta_2)^2 / (\lambda \cos\theta_2)$ and $4\pi n_3 \kappa_3 d_3 (\cos\theta_2)^2 / (\lambda \cos\theta_1)$ curves corresponding to Equation (S13). Results also show that the calculated absorptivity is higher than 92% over the entire visible spectrum validating Equations (14) and (19).

**3. V-shape solar cell structure**

Here, we preliminarily designed a viable solar cell structure. First, to efficiently absorb both TE and TM polarized light, cell 1, MoS$_2$/MgF$_2$/Ag, and cell 2, MgF$_2$/MoS$_2$/MgF$_2$/Ag, were designed to form a V-fold as shown in Figure S2. Take the case with a fold angle of 30° as an example. A 75° incident ray bounces three times on each cell and parallelly exits the V-fold (this trajectory is reversible).

Secondly, we considered the effect of reduced solar incident area caused by oblique incidence, because solar cells should be designed to generate as much electricity as possible from given cell areas, i.e. given material costs. The V-shape structure was first proposed by McGehee, Peumans and their colleagues [S3, S4]. They showed that optical absorption is further enhanced by multiple bounces on each cell, and claimed that using V-shape configuration is viable for solar cells after balancing the added cost of the shaped substrate and increase in active material use, with the reduction in installation cost for modules with higher efficiencies.

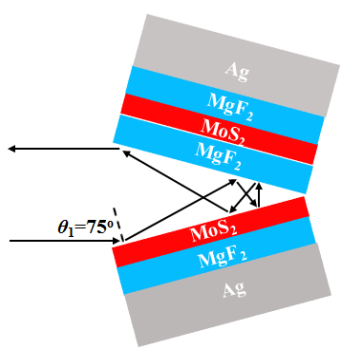

**Figure S2. V-shape solar cell structure.** A representative reversible light path is shown for 75 º incident angle.

**References**

[S1] D. Liu, Q. Li, Sub-nanometer planar solar absorber. Nano Energy 34 (2017) 172-180.

[S2] D. Liu, H.T. Yu, Z. Yang, Y.Y. Duan, Ultrathin planar broadband absorber through effective medium design. Nano Res. 9 (2016) 2354-2363.

[S3] S. Rim, S. Zhao, S.R. Scully, M.D. McGehee, P. Peumans, An effective light

trapping configuration for thin-film solar cells. Appl. Phys. Lett. 91 (2007) 243501.

[S4] S.J. Kim, G.Y. Margulis, S. Rim, M.L. Brongersma, M.D. McGehee, P. Peumans, Geometric light trapping with a V-trap for efficient organic solar cells. Opt. Express 21 (2013) A305-A312.